\documentclass[a4paper,11pt,twocolumn,pdftex]{nveart}
\usepackage{graphicx,amsmath,amssymb}

\begin{document}

\company{To appear in the proceedings of}
\journal{ICRC 2007, Merida, Mexico.}

\begin{frontmatter}
\title{GRB neutrino detection via time profile stacking \thanksref{icrc07}}
\thanks[icrc07]{Talk presented at ICRC07, Merida, Mexico.}

\author{Nick van Eijndhoven}
\address{Department of Physics and Astronomy, Utrecht University\\
         Princetonplein 5, NL-3584 CC Utrecht, The Netherlands\\
         Email : nickve.nl@gmail.com}

\begin{abstract}
A method is presented for the identification of high-energy neutrinos from
gamma ray bursts by means of a large-scale neutrino telescope.
The procedure makes use of a time profile stacking technique of observed neutrino
induced signals in correlation with satellite observations.
By selecting a rather wide time window, a possible difference between the arrival times
of the gamma and neutrino signals may also be identified. This might provide
insight in the particle production processes at the source.
By means of a toy model it will be demonstrated that a statistically significant
signal can be obtained with a km$^{3}$-scale neutrino telescope on a sample of 500
gamma ray bursts for a signal rate as low as 1 detectable neutrino for 3\% of the bursts.
\end{abstract}

\end{frontmatter}

\section{Introduction}
Cosmic radiation is a valuable source of information about various energetic astrophysical processes.
Candidates for the production of the most energetic cosmic rays are
Active Galactic Nuclei (AGN) and Gamma Ray Bursts (GRBs).

Interactions of accelerated protons and electrons with the
ambient photons at the acceleration site give rise to very energetic
secondary particles.
In particular for proton energies around the 'knee' region of the cosmic ray spectrum,
$p\gamma$ interactions yield a flux of very energetic ($>100$~TeV)
neutrinos, mainly via production of the $\Delta$ resonance and subsequent decays.
%

Various attempts\footnote{IceCube collaboration, ICRC 2005 proceedings.}
have been made to identify a high-energy
neutrino excess in correlation with satellite observations of GRBs.
The performed searches
comprise both photon-neutrino coincidence studies and investigations
of so-called "rolling time windows".
However, the former will obviously fail in case there exists a significant time difference
between the arrival times of the photon and neutrino fluxes, whereas the latter
can only be succesful in case some GRBs produce multiple neutrino detections
within the corresponding time windows.
So far, no positive identifications have been reported. 

From the above it is seen that it would be preferable to use an analysis procedure that does not
require the simultaneous arrival of photons and neutrinos and which also provides
a high sensitivity in case of low signal rates.
Such a method, based on a time profile stacking technique, is presented here and
evaluated by means of a toy model
which mimics GRB induced signals as well as (atmospheric) background.
The only large scale neutrino telescope currently in operation is
IceCube\footnote{See http://www.icecube.wisc.edu.}
and as such we use the parameters of this detector \cite{i3sens} as benchmark values
for our present studies.

\section{Signal and background generation}
Our analysis procedure is based on the detection of upgoing $\mu$ tracks and as such
our toy model only generates GRB positions homogeneously distributed over the
hemisphere opposite to the detector location.\\
For each generated burst location we define the satellite trigger time to be $t_{grb} \equiv 0$
and create a time window of $\pm 1$~hour around it.
Observations with the AMANDA neutrino telescope \cite{amaupmu} show that a km$^{3}$-scale
detector will observe on average 300 upgoing muons per 24 hours due to (atmospheric) background,
homogeneously distributed over the hemisphere and uniformly in time.
Therefore, each of the above time windows will be filled with a number of background upgoing muon
signals taken from a Poissonian distribution with an average number of 25.
To conservatively account for detector resolutions \cite{i3sens},
Gaussian spreads of $\sigma_{t}=10~\mu$s and $\sigma_{a}=1^{\circ}$ are introduced
to the arrival times and directions, respectively.

Only a fraction $f$ of the generated
burst locations is randomly selected to yield a single upgoing $\mu$ signal.
A reasonable estimate for the possible photon-neutrino arrival time difference $\tau$
and its spread $\sigma_{\tau}$ can be obtained from the actual burst duration.
Satellite observations \cite{satgrbs} exhibit a mean burst duration of about 30~seconds.
As such, the upgoing $\mu$ signal arrival time of each signal burst is taken from
a Gaussian distribution with a mean $\tau=30$~s and $\sigma_{\tau}=30$~s.
Also for these signal muon arrival times and directions the corresponding detector resolutions
$\sigma_{t}$ and $\sigma_{a}$ are introduced.

In order to optimise the time bin clustering of the signals, the bin size should be taken
to be of the order of the temporal signal spread $\sigma_{\tau}$.
However, since the observed redshifts of GRBs \cite{satgrbs} exhibit a median value of $z=1.9$
with a spread of 1.3, cosmological time dilation effects have to be taken into account.
As such, we take for the time profile bin size a conservative value of $5\sigma_{\tau}$,
corresponding to 150~s.\\
The fraction $f$ we keep as a free parameter in order to determine the sensitivity
of our analysis procedure for different sizes of the GRB sample.

\section{Analysis of simulated data}
The above results in a set of identical time windows with upgoing $\mu$ arrival time recordings
relative to the corresponding GRB trigger time.\\
Stacking of all these time profiles will exhibit a uniform distribution for
background events. However, in case the data contain upgoing $\mu$
signals correlated with the GRBs, a clustering of data bins is expected,
allowing the identification of correlated signals.
Due to the cumulative character of the procedure, large statistics can be obtained
resulting in a good sensitivity even in case of low signal rates.

\begin{figure}[htb]
\begin{center}
\includegraphics[keepaspectratio,width=7cm]{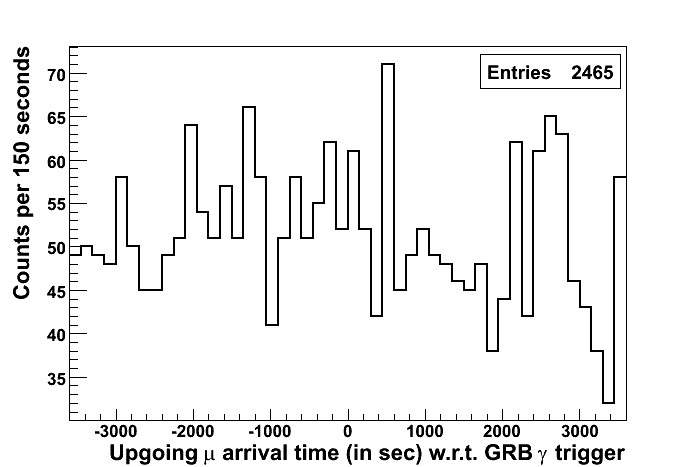}
\end{center}
\caption{Stacked time profile for 100 GRBs with $f=0.1$.
         Further details can be found in the text.}
\label{fig:tott1}
\end{figure}

For a first investigation of the performance of the procedure we generated 100 GRBs
in one hemisphere.
All parameters were set to the values mentioned above and for the fraction $f$
we used a value of 10\%\footnote{F. Halzen, {\em et al}, Astrophys. J. {\bf 527} (1999) L93.}.
The resulting stacked time profile is shown in Fig.~\ref{fig:tott1}.

\begin{figure}[htb]
\begin{center}
\includegraphics[keepaspectratio,width=7cm]{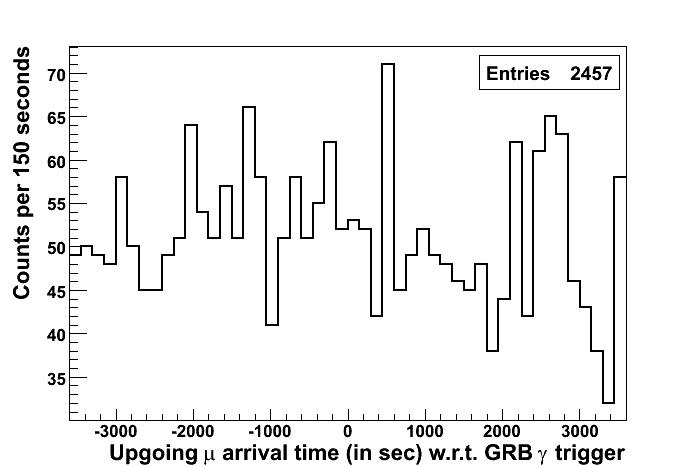}
\end{center}
\caption{Stacked time profile corresponding to the background data of Fig.~\ref{fig:tott1}.}
\label{fig:bkgt1}
\end{figure}

The corresponding stacked time profile from only the background signals
is shown in Fig.~\ref{fig:bkgt1}.

Comparison of the number of entries from Fig.~\ref{fig:tott1} and Fig.~\ref{fig:bkgt1}
shows that 8 of our generated GRBs induced a signal in the stacked time window.
However, due to the presence of a large background we are not able to identify
the GRB signals on the basis of our observations of Fig.~\ref{fig:tott1} alone.

\begin{figure}[htb]
\begin{center}
\includegraphics[keepaspectratio,width=7cm]{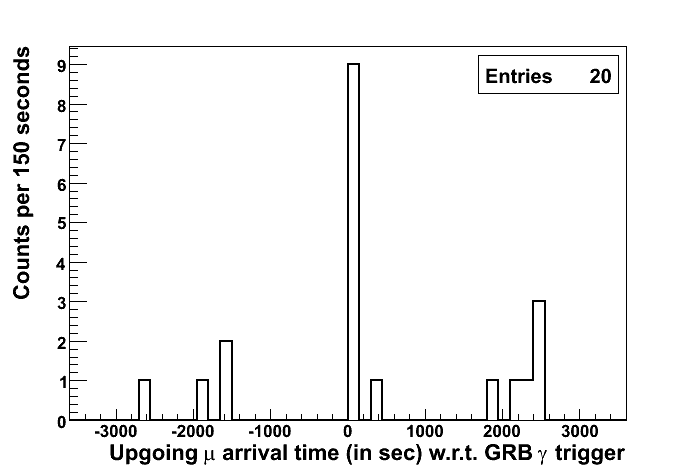}
\end{center}
\caption{Stacked time profile for 100 GRBs with $f=0.1$ and restricted to
         an angular region of $5^{\circ}$ around the actual burst location.}
\label{fig:tott2}
\end{figure}

\begin{figure}[htb]
\begin{center}
\includegraphics[keepaspectratio,width=7cm]{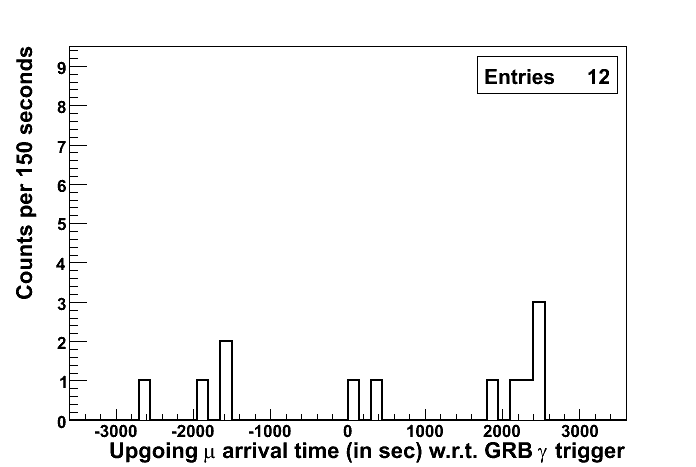}
\end{center}
\caption{Stacked time profile corresponding to the background data of Fig.~\ref{fig:tott2}.}
\label{fig:bkgt2}
\end{figure}

Restricting ourselves to an angular region of $5^{\circ}$ around the GRB location will
reduce significantly the background while preserving basically all signal muons.
The corresponding stacked time profiles of our previous generation, but now restricted to
an angular region of $5^{\circ}$ around the burst location, are shown in
Figs.~\ref{fig:tott2} and \ref{fig:bkgt2}.  

Comparison of Fig.~\ref{fig:tott2} and Fig.~\ref{fig:bkgt2} allows the identification
of the GRB signals in the central bin.
In the analysis of experimental data, however, we don't have access to the actual corresponding
background distribution. As such, we need to quantify our degree of (dis)belief in a background
observation solely based on the actually recorded signals like in Fig.~\ref{fig:tott2}.

\section{Bayesian assessment of the significance}
Consider a hypothesis $H$ and some unspecified alternative $H_{\ast}$
in the light of some observed data $D$ and prior information $I$.
We introduce the notation $p(H|DI)$ to represent the probability that $H$ is true
under the condition that both $D$ and $I$ are true.
From the theorem of Bayes we immediately obtain
\begin{equation}
\frac{p(H|DI)}{p(H_{\ast}|DI)}=\frac{p(H|I)}{p(H_{\ast}|I)}\,\frac{p(D|HI)}{p(D|H_{\ast}I)} \quad .
\label{eq:bayes2}
\end{equation}
Introducing an intuitive decibel scale, we can express the evidence $e(H|DI)$ for $H$ relative to
any alternative based on the data $D$ and prior information $I$ as~: 
\begin{equation}
e(H|DI) \equiv 10\log \left[\frac{p(H|DI)}{p(H_{\ast}|DI)} \right] \quad .
\label{eq:evidence}
\end{equation}
%
%

To quantify the degree to which the data support a certain hypothesis,
we introduce the Bayesian observables $\psi \equiv -10\log p(D|HI)$ and
$\psi_{\ast} \equiv -10\log p(D|H_{\ast}I)$.
Since the value of a probability always lies between 0 and 1, we have $\psi \geqq 0$
and $\psi_{\ast} \geqq 0$. Together with eq.~\eqref{eq:evidence} we obtain
\begin{equation}
e(H_{\ast}|DI)=e(H_{\ast}|I)+\psi-\psi_{\ast} \leqq e(H_{\ast}|I)+\psi \quad .
\label{eq:evidence3}
\end{equation}
In other words : there is no alternative to a certain hypothesis $H$ which can be supported by the
data $D$ by more than $\psi$ decibel, relative to $H$.\\
So, the value $\psi=-10\log p(D|HI)$ provides the reference to quantify our
degree of belief in $H$.

In our evaluation of the stacked time profile the main question is to which degree we
believe our observed distribution to be inconsistent with respect to a uniform background.
This question can be answered unambiguously if we are able to determine the $\psi$
value corresponding to the uniform background hypothesis $H$ based on our observed stacked time profile.

The process of recording uniform background signals is identical to performing an experiment
with $m$ different possible outcomes $\{A_{1},...,A_{m}\}$ at each trial.
Obviously, $m$ is in our case just the number of bins in the time profile, the
number of trials $n$ is the number of entries and all the probabilities $p_{k}$ corresponding
to the various outcomes $A_{k}$ on successive trials are equal to $m^{-1}$.
As such, the probability $p(n_{1} \dots n_{m}|HI)$ of observing $n_{k}$ occurrences
of each outcome $A_{k}$ after $n$ trials is given by the multinomial distribution.
%
%
This immediately yields the following expression for the corresponding $\psi$ value
\begin{equation}
\psi=-10 \left[ \log n! + \sum_{k=1}^{m}(n_{k}\log p_{k}-\log n_{k}!) \right]~.
\label{eq:psi}
\end{equation}

\section{Discovery potential}
Evaluation of the expression of eq.~\eqref{eq:psi} for the data displayed in Figs.~\ref{fig:tott1}
and \ref{fig:bkgt1} yields $\psi=713.38$~dB and $\psi_{bkg}=709.43$~dB, respectively.
Consequently, it is required to determine the $\psi$ value of the corresponding background
before the statistical significance of an observed time profile can be evaluated.

One way to investigate background signals is to record data as outlined above, but with
fictative GRB trigger times not coinciding with the actual $t_{grb}$.
In order to have similar detector conditions for both the signal and background studies,
background data may be recorded in a time span covering 1 day before and 1 day after the
GRB observation. This provides at least 25 different background time profiles
per burst, which in turn yield the corresponding different stacked background time profiles
allowing the determination of an average value $\bar{\psi}_{bkg}$ and the corrresponding
root mean square deviation $s_{bkg}$.

In the case of the situation reflected by Fig.~\ref{fig:tott1} this yields $\bar{\psi}_{bkg}=692.04$~dB
and $s_{bkg}=21.19$~dB, which is seen to be in excellent agreement with the actual background
value corresponding to Fig.~\ref{fig:bkgt1}.\\
Comparison of the actually observed $\psi$ value of 713.38~dB with the reconstructed background values
immediately shows that no significant signal is observed.\\
However, evaluation of the data corresponding to Fig.~\ref{fig:tott2} yields $\psi=218.78$~dB
with background values $\bar{\psi}_{bkg}=99.62$~dB and $s_{bkg}=23.98$~dB.
Here a statistically significant signal is obtained.

\begin{figure}[htb]
\begin{center}
\includegraphics[keepaspectratio,width=7cm]{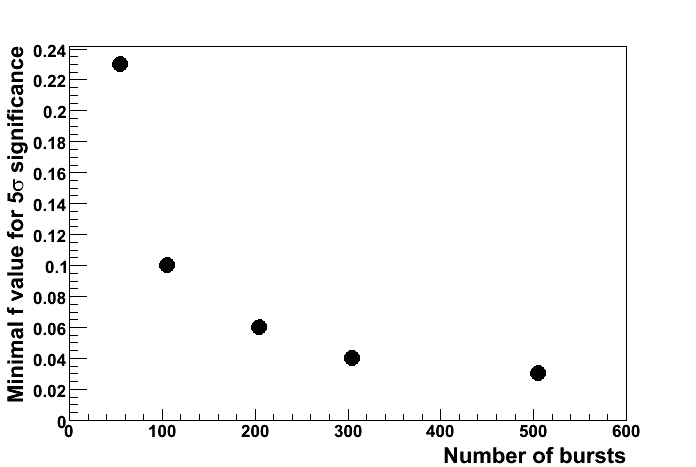}
\end{center}
\caption{Sensitivities corresponding to a $5\sigma$ signal significance.}
\label{fig:disc}
\end{figure}

Variation of the number of GRBs allows a determination of the minimal value of the fraction $f$
for which a statistically significant signal can be obtained.
Common practice is to claim a discovery in the case a significance in excess of $5\sigma$
is obtained. Following the procedure outlined above this leads to the discovery sensitivities
as shown in Fig.~\ref{fig:disc}.

%

\section{Summary}
The method introduced in this report allows identification of high-energy neutrinos
from gamma ray bursts with large scale neutrino telescopes.
The procedure is based on a time profile stacking technique, which provides statistical
significant results even in the case of low signal rates.

The performance of the method has been investigated by means of toy model studies based on
realistic parameters for the future IceCube km$^{3}$ neutrino telescope and a variety of burst samples.
From these investigations it is seen that a $5\sigma$ significance is obtained on a
sample of 500 bursts with a signal rate as low as 1 detectable neutrino for 3\% of the bursts.\\


\end{document}